\documentclass[%
reprint,
 amsmath,amssymb,
 aps,
prstab,
floatfix,
]{revtex4-2}

\usepackage{graphicx}
\usepackage{dcolumn}
\usepackage{bm}
\usepackage{caption}
\usepackage{subcaption}
\usepackage{ulem}

\begin{document}

\preprint{APS/123-QED}

\title{Modeling and Design of Compact, Permanent-Magnet Transport Systems for Highly Divergent, Broad Energy Spread Laser-Driven Proton Beams}

\author{J.~De~Chant}
\altaffiliation[Also at ]{Physics Department, Michigan State University.}
\author{K.~Nakamura}%
\email{KNakamura@lbl.gov}
\author{Q.~Ji}%
\author{L.~Obst-Huebl}%
\author{S.~Barber}%
\author{A.~M.~Snijders}%
\author{C.~G.~R.~Geddes}%
\author{J.~van Tilborg}%
\author{A.~J.~Gonsalves}
\author{C.~B.~Schroeder}%
\author{E.~Esarey}%

\affiliation{%
 Lawrence Berkeley National Laboratory, 1 Cyclotron Road, Berkeley, California 94720, USA
}%


\begin{abstract}
Laser-driven (LD) ion acceleration has been explored in a newly constructed short focal length beamline at the BELLA petawatt facility (interaction point 2, iP2). For applications utilizing such LD ion beams, a beam transport system is required, which for reasons of compactness be ideally contained within 3 m. While they are generated from a micron-scale source, large divergence and energy spread of LD ion beams present a unique challenge to transporting them compared to beams from conventional accelerators. This study gives an overview of proposed compact transport designs using permanent magnets satisfying different requirements depending on the application for the iP2 proton beamline such as radiation biology, material science, and high energy density science. These designs are optimized for different parameters such as energy spread and peak proton density according to the application’s need. The various designs consist solely of permanent magnet elements, which can provide high magnetic field gradients on a small footprint. While the field strengths are fixed, we have shown that the beam size is able to be tuned effectively by varying the placement of the magnets. The performance of each design was evaluated based on high order particle tracking simulations of typical LD proton beams. We also examine the ability of certain configurations to tune and select beam energies, critical for specific applications. A more detailed investigation was carried out for a design to deliver 10 MeV LD accelerated ions for radiation biology applications. With these transport system designs, the iP2 beamline is ready to house various application experiments.

\end{abstract}

\maketitle

\section{\label{sec:level1}Introduction}
Laser-driven acceleration has become a very attractive area of research due to the unique properties of the ion beams generated. In a laser-driven ion accelerator, a high intensity laser pulse is focused onto a thin target, resulting in the acceleration of a large number of protons (and/or heavier ions) to high energies out of the target. Compared to conventional radiofrequency (RF) accelerators, laser-driven ion beams exhibit unique properties, including a large number of protons per shot (10$^{13}$/shot), a low emittance ($< 0.1$~mm~mrad), and an ultra-short pulse duration (10$^{-12}$~s at the source) \cite{Macchi_2017,daido_2012}.
A significant advantage of laser-ion accelerators is their large accelerating gradient which is four orders of magnitude greater than in a traditional RF cavity (10$^6$~MV/m vs. 10$^2$~MV/m).
The large accelerating gradient allows ions to be accelerated to a higher energy over a shorter distance. This property makes such accelerators advantageous for applications that need to be compact and can provide broader access for users of research accelerators as they can be set up in smaller labs. The ultra-short pulse duration and high energy density of laser-driven ion beams also makes them suitable for ultra high dose-rate radiobiological studies related to cancer therapy~\cite{Ma_2006, Bin_2022, Kroll_2022}. Other applications can be found in the fields of material science~\cite{Redjem_2023} and warm dense matter physics~\cite{Patel_2003, Antici_2006, Malko_2022}.

In order to take full advantage of the acceleration mechanism's compact nature, it is preferable for the beam transport to also be compact. This has proven to be a difficult task however, as LD ion beams are characterized by an initially large divergence and energy spread at the source. To accommodate such beams, transport systems using high gradient focusing elements have been utilized in the field, typically via pulsed solenoids~\cite{Burris-Mog_2011,Jahn_2018,Brack_2020} or permanent magnet quadrupoles (PMQs)~\cite{Lim_2005,Eichner_2007}. The beam transports based on PMQs have been demonstrated for up to 14 MeV laser-driven proton beams~\cite{Ter-Avetisyan_2008,Schollmeier_2008,Nishiuchi_2009,Rosch_2020}. Another option is to use an active plasma lens (APL)~\cite{Panofsky_1950, vanTilborg_2015, Lindstrom_2018, Bin_2022}. For certain applications, APLs can be attractive due to the tunability of the focusing force and azimuthal symmetry though the radial acceptance of APLs is not readily expandable which limits its collection efficiency. Superconducting magnets~\cite{Wan_2015} can achieve a very strong field gradient with a large bore but, not only can they become prohibitively expensive, they require a large cryogenic apparatus which exceeds the size requirements. They also run the risk of quenching from being so close to the laser-plasma interaction. Lastly, if compactness is not required, electromagnets and hybrid designs with technologies mentioned above can provide extended capabilities~\cite{Romano_2016}. 

At the Lawrence Berkeley National Laboratory (LBNL), Berkeley Lab Laser Accelerator (BELLA) Center, laser-driven acceleration of proton beams has been pursued using its existing large f-number beamline~\cite{Nakamura_2017, Steinke_2020}. Recently, an additional short focal length, high intensity laser beamline to deliver higher laser intensities to the target, called interaction point 2 (iP2), has been constructed and is available to users through LaserNetUS \cite{Toth_2017,Hakimi_2022,ObstHuebl_2023}. For this new beamline, a compact high-energy proton beam transport system is essential for applications of the laser-accelerated proton beams. 

The requirements for the beam transport depend on these applications. For example, radiobiology studies require an adjustable beam size to match the lateral extent of the sample, while isochoric heating may benefit from maximized beam fluence in a small focus on the sample. Furthermore, the ion beam transport for this new beamline is not only preferred to be compact, but is required to be so due to its limited real estate surrounding the iP2 target chamber. The total length of transport, including beam optics, diagnostics, and ion sample, must be confined to 3~m.

In this paper, we report on a design study for such compact ion beam transports, transporting up to 30~MeV protons to a dedicated sample site. We chose to focus on beam optical elements using permanent magnets considering the cost and space efficiency, relatively high field gradient, and their established performance under harsh laser-plasma environments~\cite{Ter-Avetisyan_2008, Schollmeier_2008, Nishiuchi_2009,Rosch_2020}. Permanent magnets can be arranged in a Halbach arrangement to boost the overall strength given the limited remanent magnetization of commercially available rare-earth ferromagnetic materials~\cite{Halbach_1980}.Various configurations were explored and their performances were analyzed so that users can choose the optimized configuration for their applications. 

This work lays the foundation for the broader application of laser-driven ion beams by addressing two of their most significant challenges: large divergence and broadband energy spread. The tools and designs presented here will enable current and future laser-driven ion accelerator facilities to more effectively meet the diverse needs of users across various applications. As an example, we present a beamline system specifically designed to collimate 10 MeV protons for radiobiological studies at iP2. This system, utilizing only two permanent magnets, has already been successfully implemented and used in radiobiological experiments \cite{dechant_2024}.

\section{Methods\label{sec:Methods}}
\subsection{Design Tools}

Particle optics simulation tools are typically divided into two categories: numerical field integrators and map codes. Numerical field integrators, such as TRACK~\cite{TRACK}, IMPACT~\cite{IMPACT}, G4Beamline~\cite{G4Beamline}, and BDSIM~\cite{BDSIM},  model the complete electromagnetic fields in a system and simulate the trajectories of individual particles by integrating the equation of motion. Numerical field integrators are robust in their description of the beam dynamics but can be computationally expensive, making them less suitable for rapid lattice optimization and simulating large particle numbers. 
Conversely, map codes like MADX~\cite{MADX}, TRANSPORT~\cite{TRANSPORT}, and TRACE-3D~\cite{TRACE3D}, model the transport system as transfer matrices to describe the action of each optical element on the beam in phase space. This approach is favored for optimization tasks due to its speed and the ability to quickly analyze optical properties of the system represented by the elements of the transfer matrix.  This enables quick visualization and aberration correction, making them highly advantageous for streamlining the design process and enhancing system performance.

In this work, we utilised COSY INFINITY~\cite{COSY_INFINITY}, which combines the speed of traditional map codes with the precision of numerical field integrators. COSY is commonly used in the the study of accelerator lattices, spectrographs, beam transports, electron microscopes, and many other devices. Using differential algebraic techniques to perform the numerical integration of the fields, it calculates Taylor expansions of the transfer matrix to arbitrarily high order. 


COSY outputs the map of interest, $M$, relating its final phase-space vector $\vec{z_f}$ to its initial phase-space vector $\vec{z_0}$ and some other parameters $\vec{\delta}$ (including charge, mass, and energy),
\begin{equation}
    \vec{z_f}=M(\vec{z_0},\vec{\delta}).
\end{equation}
Each element in $M$ represents coupling between elements in the particle's state vector. For example, $M_{12}=(x|x')$ ($M_{34}=(y|y')$) represents the coupling between a particle's position in the $x$ ($y$) plane and the particle's transverse momentum $x'$ ($y'$). Given a configuration of magnets or electrostatic fields, COSY INFINITY can approximate a transfer map to high order and optimize magnet strengths and positions to achieve focusing. The transfer map needs to be known to high order here as non-linear effects are important due to the large beam size and energy spread of LD ion beams. The trajectories of a collection of particles can then be calculated by applying the map to the particles' initial values of position and momentum. COSY provides particle tracking for a small number of particles and can plot their trajectory through the transport. The COSY output maps were used to evaluate the different investigated transport systems' angular acceptance and transmission efficiency. In order to simulate and analyze the performance of the transport systems with a large number of particles, a custom MATLAB script was written.


\subsection{Input beam parameters}\label{subsec:beam_props}
In order to estimate and compare the performance of each transport design, a beam with initial parameters typical of laser driven proton accelerators in the target normal sheath acceleration (TNSA) regime~\cite{Macchi_2017} was generated. The maximum laser intensity in the focus for the BELLA iP2 is estimated at $\approx5\times10^{21}$~W/cm$^2$. For this beamline, the TNSA energy spectrum for the accelerated proton beam was modeled as, 
\begin{equation}\label{eq:boltz}
    N(E)=5 \times 10^{10} e^{-E/(7 \text{ MeV})},
\end{equation}
where $N$ is the number of protons in a beam and $E$ the kinetic energy of the protons.

The initial position and propagation angle of each particle were assigned corresponding to a Gaussian distribution with a FWHM of 0.1~mm of source size and 450~mrad of full beam divergence, respectively. The initial angular divergence in real laser plasma accelerated beams is a function of the particle's energy~\cite{Bin_2013} though here it is treated as constant. Particles with an initial divergence larger than the angular acceptance of the first magnet are neglected for the rest of the calculation but are used to estimate the collection percentage. After the initial beam is prepared, each particle's phase-space vector is passed through the map which calculates each particle's final phase-space vector. The performance of each transport design was then evaluated by calculating the peak density, energy acceptance, and beam spot size of the output beam. 

\subsection{Permanent Magnet Quadrupoles}

The magnetostatic simulation code RADIA~\cite{Chubar_1998} was used to design the PMQs. The simulated field distribution of the magnets in RADIA were not used in any beam transport simulations but were done only to determine the feasibility of manufacturing magnets with the desired field strength.

The magnets were based on an 8-azimuthal-segment Halbach array with 1.29~T of the bulk magnetisation $B_r$. This configuration is readily achievable for commercial vendors, and this bulk magnetisation level is typical for a N40 grade NdFeB magnet, which is also commercially available at relatively low cost. A list of the properties of the PMQs used in this design work is shown in Table \ref{tab:table1}. Since the main purpose of this work is to explore different configurations of the transport systems, only the magnet length was varied to change the focusing strength of the PMQs for simplicity, while keeping the other parameters, such as inner radius and the magnetic field at the tip, fixed to the parameters shown in the Table \ref{tab:table1}.

\begin{table}[!tb]
\caption{PMQ fixed parameters used for transport design. The length, number, and distance between the magnets were varied}
\label{tab:table1}
\begin{ruledtabular}
\begin{tabular}{lr}
Parameters & Value\\
\hline
Inner radius & 10~mm\\
Outer radius & 30~mm\\
Bulk magnetisation $B_r$ & 1.29~T\\
Effective magnetic field at tip $B_{tip}$ & 1.16~T\\
Target - 1st PMQ drift space $D_1$ & 40~mm\\
Minimum magnet - magnet distance & 80~mm\\
Minimum last magnet - output distance & 40~mm\\
\end{tabular}
\end{ruledtabular}
\end{table}

\begin{figure}[!b]
\includegraphics[width = .4\textwidth]{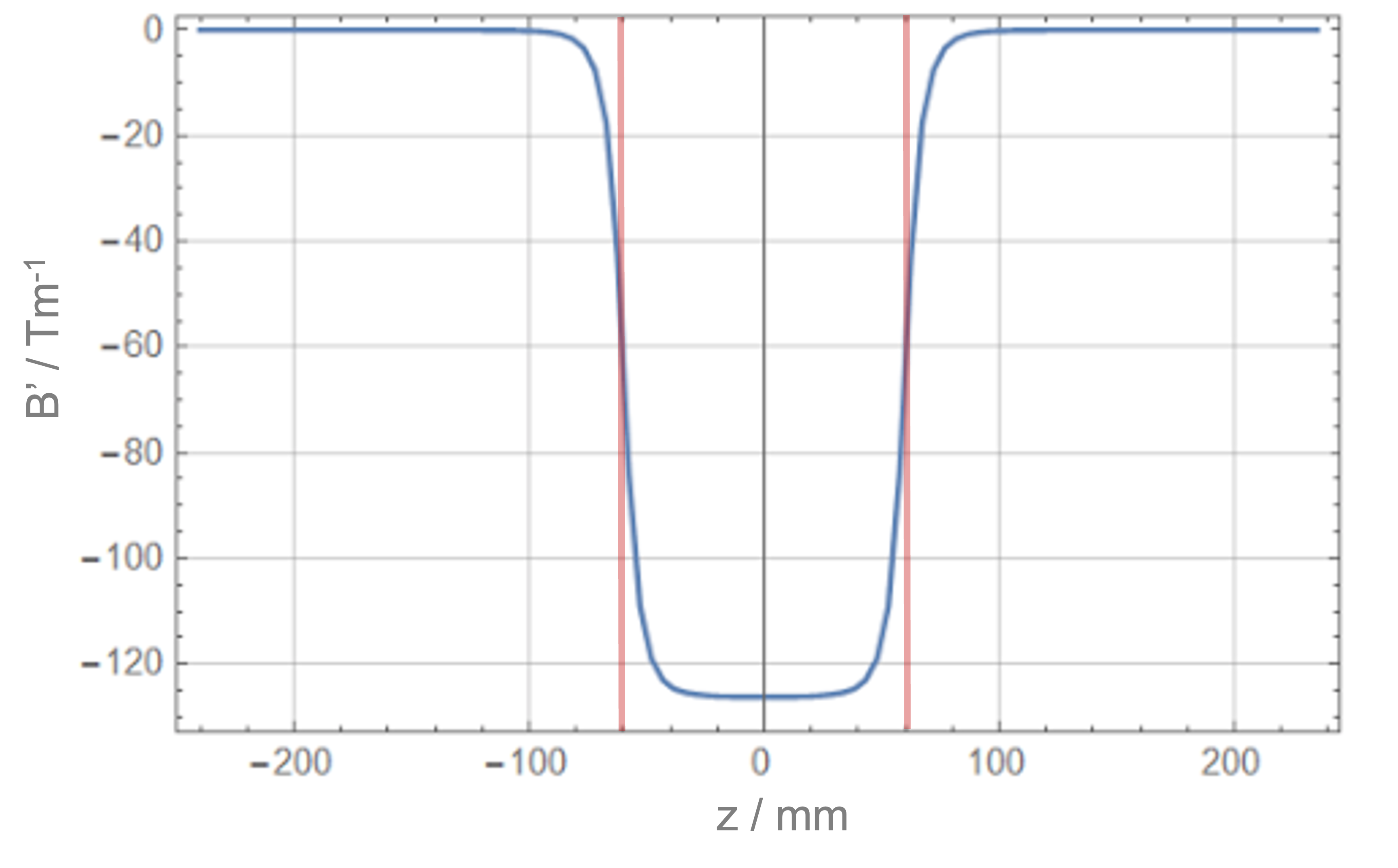}
\caption{\label{fig:quad field} Simulated field gradient as function of the propagation axis $z$. The physical length is 120~mm (shown in red) and other parameters were described in Table~\ref{tab:table1}.}
\end{figure}

The simulated field gradient for a 120~mm PMQ along the propagation axis $z$ is shown in Fig.~\ref{fig:quad field}, where the maximum field gradient was 126.2~T/m with the magnet effective length of 120.2~mm giving $B_{tip} = 1.26$~T. The effective magnetic field at the surface of one of the magnet segments, $B_{tip}$, depends on the bulk magnetisation, the magnet outer radius and the length. It was found that $B_{tip} = 1.16$~T was readily achievable for any length of the PMQs, and therefore used in this study. 

The magnetic fields in the COSY simulation are based on the built-in magnet distributions in COSY. One can see from Fig.~\ref{fig:quad field} that the fringe fields taper off $(>99\%$) about 40~mm from the physical edges of the magnet. Thus in COSY, the fringe fields on either side of each magnet were modeled using the most sophisticated approximation available in COSY which models them as falling according to the ENGE function outlined in~\cite{COSY_INFINITY}. The minimum drift length between the PMQs was set to 80~mm and between the last magnet and system output was set to 40~mm to avoid overlap of the fringe fields, which can reduce effective strength of the PMQs. The drift space between the target and the first PMQ $D_1$ was fixed to 40~mm for the same reason in addition to the beam collection efficiency being of primary concern.  Between these values, the magnet and drift lengths were varied to find the optimal magnet locations to achieve focusing at the sample plane.

\subsection{Permanent Magnet Dipoles}\label{sec:PMD}
\begin{figure}[!b]
\includegraphics[width = .4\textwidth]{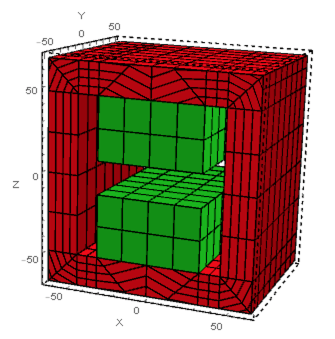}
\caption{\label{fig:pmd} Illustration of the permanent magnet dipole. The yoke is indicated in red and the magnets are indicated in green. The segmentation is for simulation purposes.}
\end{figure}

Depending on requirements from applications, dipole magnets may be desired to manipulate the energy spectrum of the beam after transport. Dipole magnets were considered for a number of transport designs discussed in this study, at different locations along the beamline. They were also designed using RADIA. They consist of a rectangular shape yoke made of 1018 steel and NdFeB permanent magnet blocks as illustrated in Fig.~\ref{fig:pmd}. The bulk magnetization was assumed to be $b_r = 1.29$~T, identical to the PMQs discussed above.

The geometrical parameters of permanent magnet dipoles (PMDs) are summarized in Table~\ref{tab:pmd}. The gap of the dipole magnet was set to be 20~mm to match the inner radius of the PMQ. The magnet length was 50.8~mm to ensure flat magnetic field profile ($<3\%$) within 20~mm horizontally. The longitudinal length was varied depending on different applications. RADIA simulations showed that the fringe field extends 40~mm on each side, therefore the same 80~mm minimum magnet to magnet distance was imposed to the design work. 

\begin{table}[!t]
\caption{PMD parameters\label{tab:pmd}}
\begin{ruledtabular}
\begin{tabular}{lr}
Parameters & Value\\
\hline
Half gap & 10~mm\\
Height & 124.8~mm\\
Bulk magnetisation $B_r$ & 1.29~T\\
Effective magnetic field & 0.89~T\\
Minimum magnet - magnet distance & 80~mm\\
\end{tabular}
\end{ruledtabular}
\end{table}

\section{Beam transport design results\label{sec:results}}

\subsection{Configuration design and performance}
    \subsubsection{Doublet\label{sbsec:dblt}}

\begin{figure}[!b]
\includegraphics[width = .45\textwidth]{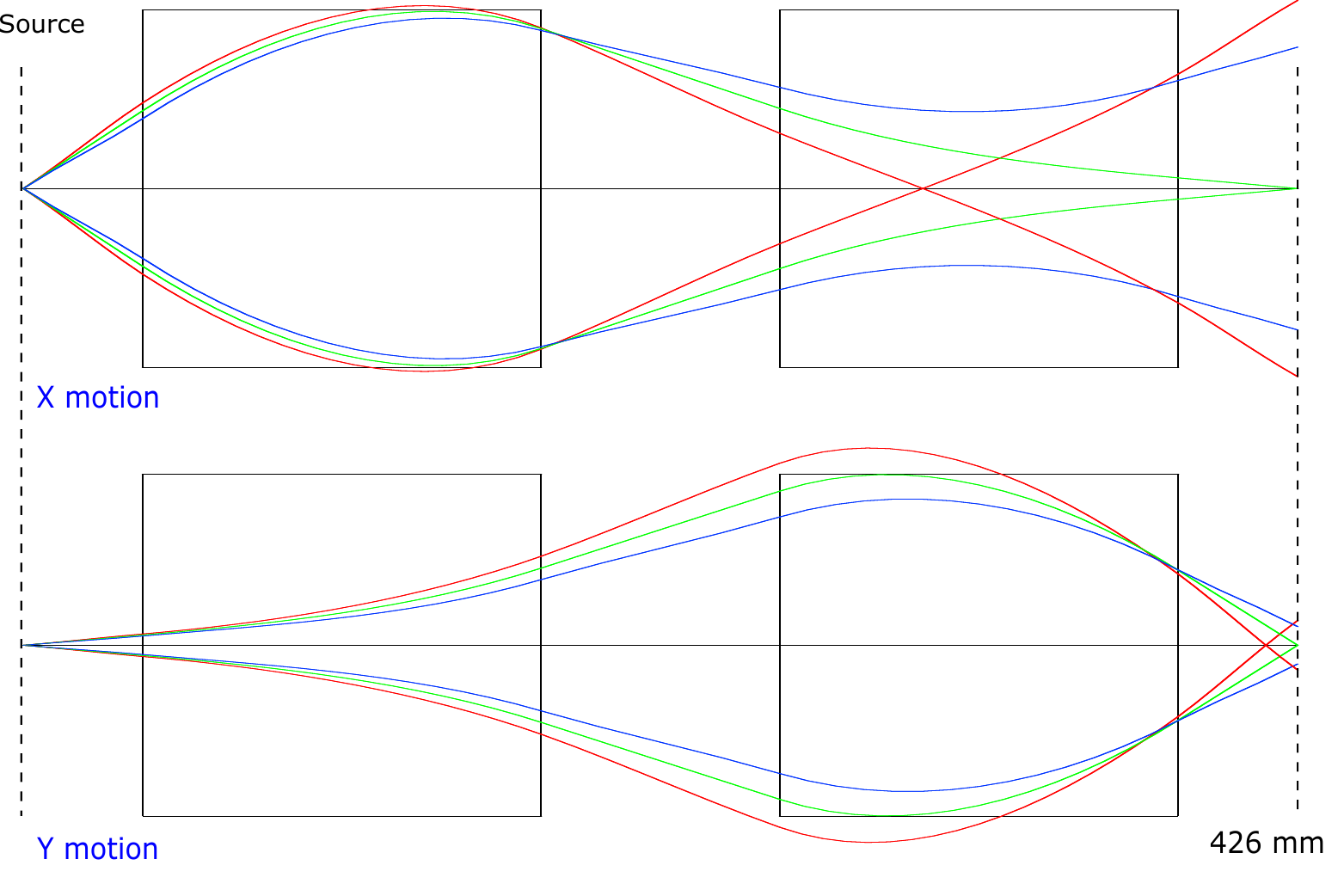}
\caption{\label{fig:dblt} Illustration of the 430~mm doublet for $E_0 = 30$~MeV proton beam. The red, green and blue lines show the trajectories for proton beams with energies of $0.8~E_0, E_0$, and $1.2~E_0$ with the maximum accepted divergence in each plane ($x' = 109$~mrad and $y' = 15$~mrad for $E_0 = 30$~MeV)}
\end{figure}

The doublet consists of two PMQs, each providing focusing force in either the x- or the y-plane, and is a widely used configuration~\cite{Ter-Avetisyan_2008,Schollmeier_2008,Nishiuchi_2009,Rosch_2020}. The system can be configured to achieve point-to-point imaging in both transverse planes ($M_{12}=M_{34}=0$). As an example, the shortest doublet system can be designed by having all the drift lengths minimum. With two variables $L_1$ and $L_2$ to meet two conditions $M_{12} = M_{34}=0$, the shortest (430~mm from source-to-focus) 30~MeV doublet was designed and illustrated in Fig.~\ref{fig:dblt}, where the beam envelope for the $E_0 = 30$~MeV beam is shown by the green lines, and the beam envelopes for the kinetic energies $0.8~E_0$ and $1.2~E_0$ are shown by the red and blue lines, respectively. The collection efficiency of the focusing system $\eta$ is defined by the angular acceptance of the $E_0 = 30$~MeV proton beams, though it is more complicated particles with different energy would have a different acceptance, as illustrated in Fig.~\ref{fig:dblt}. The angular acceptance angles $\theta_x$ and $\theta_y$ are listed in Table~\ref{tab:SysPrm}.

One can see from Fig.~\ref{fig:dblt} that the beam angular acceptance is limited by the first PMQ for the x plane, and by the second PMQ for the y plane. Since the second magnet is further downstream, and the beam gets further defocused by the first, angular acceptance is less in the y plane than the x plane. This is why in the final design, discussed in \ref{sec:10MeVCol}, the bore of the second magnet was enlarged to improve the overall collection efficiency. In the exploratory part of this paper, the bore sizes were fixed in order to compare different configurations.

\begin{table*}[!t]
\caption{\label{tab:SysPrm} Parameters of the systems, units are in mm for drift (D) and magnet lengths (L) and mrad for angular acceptance $\theta$. Some designs were optimized to deliver beams of different nominal energies (i.e. Doublet @ 210 MeV)}
\begin{ruledtabular}
\begin{tabular}{ccccccccccccccccc}
Name & Total Length & $\theta_x$ & $\theta_y$ & $D_1$ & $L_1$ & $D_2$ & $L_2$ & $D_3$ & $L_3$ & $D_4$ & $L_4$ & $D_5$ & $L_5$ & $D_6$ & $L_6$ & $D_7$\\ 
\hline
Doublet & 430 & 109 & 15 & 40 & 133.2 & 80 & 133.2 & 40\\
Doublet @ 210 MeV & 1000 & 48 & 6.2 & 157.7 & 133.2 & 412 & 133.2 & 157.7\\
Doublet & 750 & 109 & 15.9 & 40 & 126.3 & 80 & 55.3 & 448.3\\
Doublet & 1000 & 109 & 16.3 & 40 & 124.2 & 80 & 49.9 & 705.9\\
Triplet & 560 & 110 & 17.5 & 40 & 118 & 80 & 79.5 & 80 & 118 & 40\\
Triplet @ 117 MeV & 1000 & 48 & 9.5 & 97.5 & 118 & 282.5 & 79.5 & 170.4 & 118 & 131.3\\
Triplet & 1000 & 46.5 & 23.5 & 40 & 95.5 & 80 & 69.9 & 80 & 43.6 & 591.1\\
Quartet & 610 & 52 & 21.5 & 40 & 102.4 & 80 & 77.1 & 80 & 76.1 & 80 & 33.8 & 40\\
Quartet @ 92 MeV & 1000 & 24 & 14.5 & 59.6 & 102.4 & 225.3 & 77.1 & 133.8 & 76.1 & 198.2 & 33.8 & 132.4\\
Quartet & 1000 & 28.5 & 32.5 & 40 & 66.6 & 80 & 67.9 & 80 & 60.6 & 80 & 27.2 & 496\\
Quartet+Dipole & 1000 & 28.5 & 32.5 & 40 & 66.4 & 80 & 68 & 80 & 60.3 & 80 & 26.5 &  80 & 101.6 & 175.7 & 101.6 & 40\\
\end{tabular}
\end{ruledtabular}
\end{table*}

\begin{figure}[!t]
\includegraphics[width = .4\textwidth]{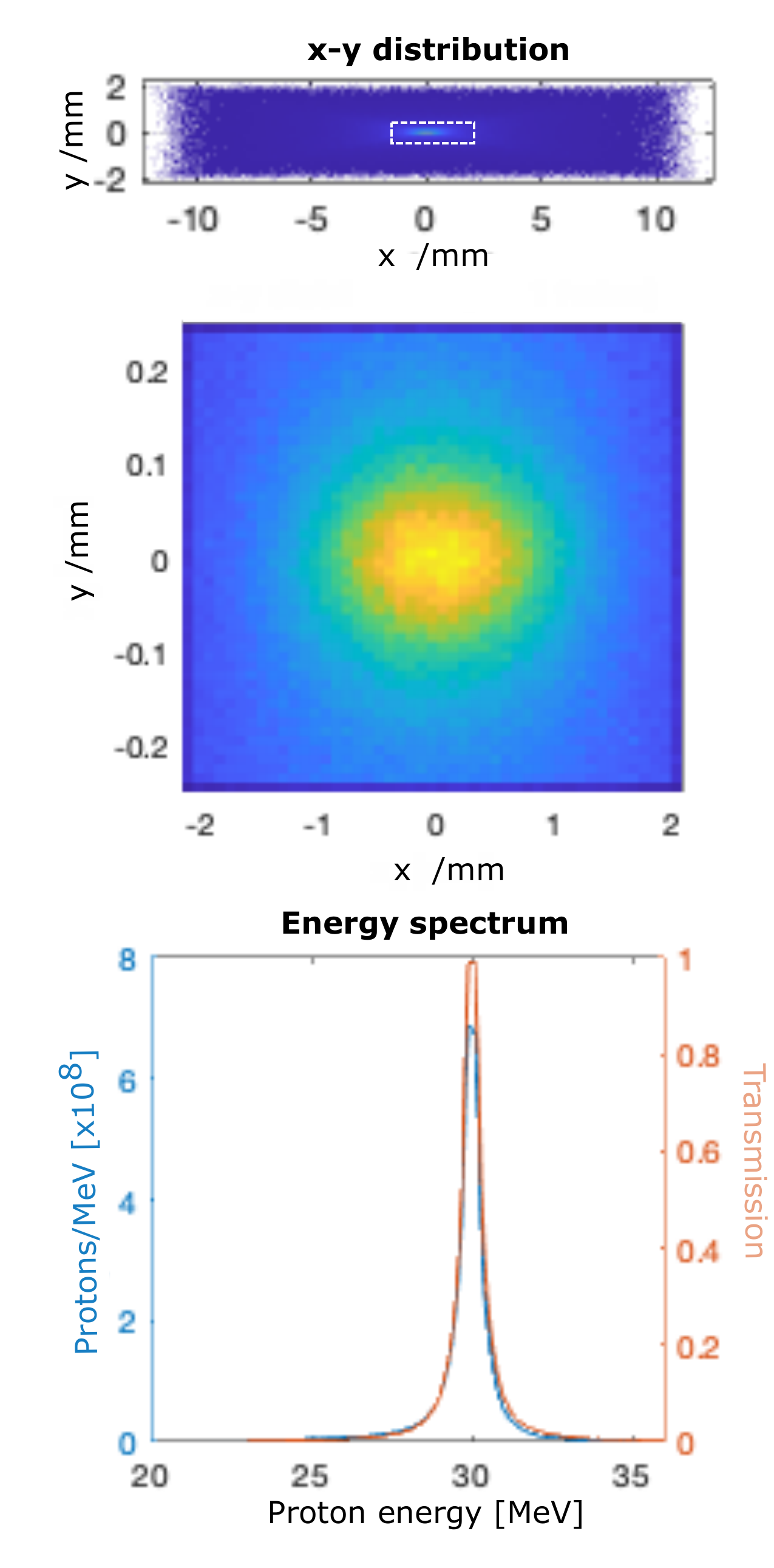}
\caption{\label{fig:roi} The output transverse beam profile of the 430~mm doublet (top, middle) and the output beam energy spectrum and energy-dependent normalized transmission through the beam line, within the limited transverse beam profile area of $\pm 1$~FWHM indicated by the dashed white line in the top image.}
\end{figure}

Shown in Fig.~\ref{fig:roi} (top) is the transverse beam profile at the focal position on the right of Fig.~\ref{fig:dblt} for the 430~mm doublet output and its energy spectrum (bottom). One can see that the output transverse beam profile for the doublet was quite asymmetric, with different levels of magnification in the x and y planes. Here the energy spectrum was evaluated by limiting the spatial integration of the beam particles to within one FWHM of the output spot size. This is equivalent to limiting the beam going into the spectrometer with an iris in the laboratory. 

The performance parameters of the 430~mm doublet, namely the collection efficiency $\eta$, the peak density of the proton beam $n_{1pk}~[10^9$ particles/mm$^2$] at the output plane, the energy spread in FWHM $\delta E$~[MeV], and output beam size in FWHM $\sigma$~[mm], are shown in Table~\ref{tab:length}. This configuration would capture 2.7\% of 30~MeV protons generated at the source can be delivered to the output of the transport for the source parameters chosen for this study. In reality, the overall acceptance of a given beamline depends on the real source parameters and may be higher due to the energy-dependent divergence of LD ion beams~\cite{Macchi_2017}.

\begin{table}[!tb]
\caption{\label{tab:length} Simulated beam transport results for doublet configurations of different lengths, including the collection efficiency $\eta$, peak ion density $n_{1pk}$, energy acceptance $\delta E$, and FWHM beam sizes $\sigma_x$, $\sigma_y$ at the output plane.}
\begin{ruledtabular}
\begin{tabular}{c|ccccc}
Length  & $\eta$ & $n_{1pk}$ & $\delta E$& $\sigma_{x}$& $\sigma_{y}$\\ 
~[mm] & [$\%$] & [10$^9$/mm$^2$] & [MeV] & [mm] & [mm] \\ 
\hline
430 & 2.70 & 16.7 & 0.65 & 0.76 & 0.034 \\
750 & 2.86 & 3.4 & 1.15 & 1.5 & 0.161 \\
1000 & 2.93 & 1.8 & 1.2 & 1.98 & 0.23 \\
\end{tabular}
\end{ruledtabular}
\end{table}

One can design doublet systems of various sizes, and the solution may not be unique. For doublet designs, one can maximize the angular acceptance of the beam by minimizing the drift lengths, $D_1$ and $D_2$. Also shown in Table~\ref{tab:length} are performance parameters for the 30~MeV 750~mm and 1.0~m doublet systems with the minimum $D_1$ and $D_2$, whose system parameters are shown in Table~\ref{tab:SysPrm}. The shorter system delivers higher proton density with a narrower energy spread and smaller spot size. The longer system can provide larger beams with small improvement in the collection efficiency and focus symmetry. When an application demands high proton density, one might employ the shortest transport system that can be achieved from a technical implementation standpoint.

Once manufactured, the magnet strength cannot be varied for PMQs. Therefore, their locations have to be varied to focus particles with different energies. The mentioned 430~mm doublet was designed for 30~MeV, and could not focus lower energy protons because all the drift lengths were minimum. The same set of PMQs used in the 430~mm doublet configuration can be rearranged within a 1~m total system length to focus proton beams with nominal energy up to 210~MeV. The system parameters for the rearranged 430~mm doublet configuration to focus a 210~MeV proton beam (now 1000 mm in length) are also listed in Table~\ref{tab:SysPrm} as "Doublet @ 210 MeV". These were estimated using the same beam described in Sec.\ref{subsec:beam_props} but with a nominal energy of 210 MeV. Note that this comes with significant reduction of the angular acceptance since the first PMQ needs to be moved downstream, but shows some flexibility of the transport even after the magnets are constructed. 

\subsubsection{Triplet configuration improves beam size symmetry}

\begin{figure}[!b]
\includegraphics[width = .45\textwidth]{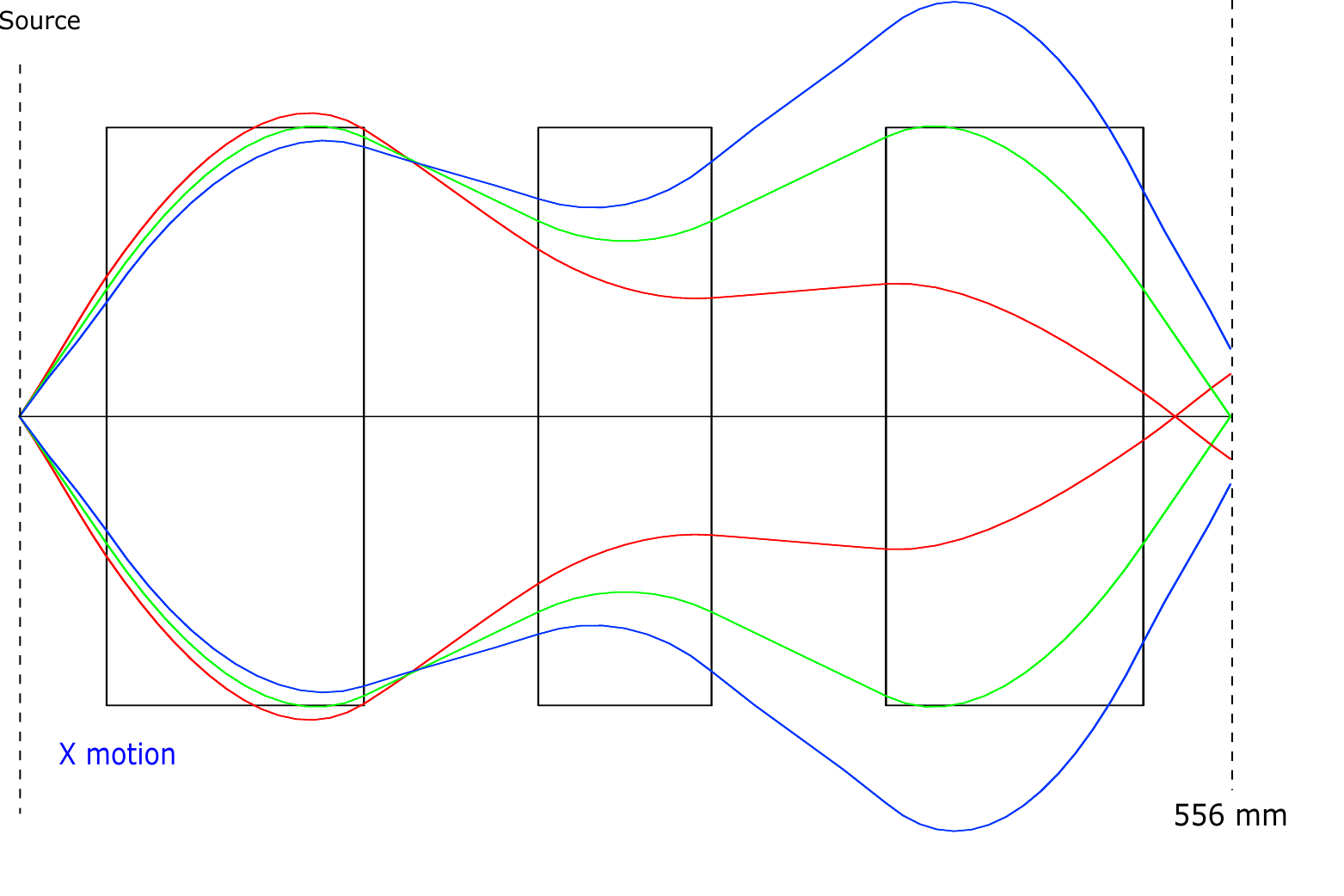}
\caption{\label{fig:trplt} Illustration of the 560~mm triplet for $E_0 = 30$~MeV proton beam. The red, green and blue lines show the envelopes of $0.8~E_0, E_0$, and $1.2~E_0$ proton beams. The half angle of the traces are 110~mrad for x and 17.5~mrad for y planes.}
\end{figure}

The beam asymmetry can be improved by requiring the $x$ and $y$ magnification to be symmetric in the design (i.e. requiring $M_{11}=M_{33}$). This can be accomplished by adding another focusing magnet and thus another free parameter. The shortest triplet design with symmetric magnification ($M_{12} = M_{34} = 0, M_{11} = M_{33}$) and a system length of 560~mm is illustrated in Fig.~\ref{fig:trplt}. For this design, all drift lengths were fixed to be the minimum length and all PMQ lengths $L_1, L_2$ and $L_3$ were treated as variables.  

The system parameters and performance parameters for 0.56~m and 1.0~m triplets are shown in Table~\ref{tab:SysPrm} and Table~\ref{tab:cnfg}, respectively. While the solution for the 1.0~m triplet is not unique, $D_2$ and $D_3$ were kept minimum for comparison with the doublet performance. By comparing the shortest configurations, the triplet delivers the higher proton peak density with much improved symmetry compared to the doublet, while the efficiency and the energy spread were comparable. When comparing the systems with the same length of 1~m, the triplet delivers again higher proton peak density with improved symmetry despite of significantly lower collection efficiency than the doublet. The energy spread was comparable. It is clear that some asymmetry still remains for the triplet designs. This is due to the asymmetry in the angular acceptance and chromatic aberrations. With insignificant cost to the available space, the triplet design would deliver better beam quality than the doublet for the applications envisioned at iP2.

\begin{table*}[!tb]
\caption{\label{tab:cnfg} Simulated beam transport results for different transport configurations, including the collection efficiency $\eta$, peak proton density $n_{1pk}$, energy acceptance $\delta E$, and FWHM beam sizes $\sigma_x$, $\sigma_y$ at the output plane. The minimum achievable transport map coefficients are provided to highlight a given configuration's symmetry with values closer to 1 in the $x$ ($M_{11}$) and $y$ ($M_{33}$) planes indicating point-to-point focusing and similar values indicating similar magnification.}
\begin{ruledtabular}
\begin{tabular}{ccccccccc}
Name & Length & $\eta$ & $n_{1pk}$ & $\delta E$ & $\sigma_{x1}$ & $\sigma_{y1}$  &$M_{11}$ & $M_{33}$\\
&~[mm] & [$\%$] & [10$^9$/mm$^2$] & [MeV] & [mm] & [mm] \\ 
\hline
Doublet & 430 & 2.2 & 18 & 0.75 & 0.84 & 0.034 & 7.3 & 0.14 \\
Triplet & 560 & 2.5 & 25 & 0.85 & 0.13 & 0.17 & 1.0 & 1.0 \\
Quartet & 610 & 1.5 & 24 & 0.85 & 0.15 & 0.19 & 1.1 & 1.1 \\
\hline
Doublet & 1000 & 2.3 & 2.0 & 1.4 & 2.1 & 0.23 & 17 & 1.2 \\
Triplet & 1000 & 1.5 & 5.1 & 1.4 & 0.36 & 0.53 & 2.8 & 2.8 \\
Quartet & 1000 & 1.3 & 6.6 & 1.2 & 0.34 & 0.35 & 2.4 & 2.4 \\
\hline
Quartet (z=-200 mm) & 800 & 1.3 & 11 & 1.7 & 0.38 & 0.48 \\
Quartet (z=500 mm) & 1400 & 1.3 & 2.0 & 2.4 & 0.69 & 0.69 \\
\hline
Mirrored doublet & 2000 & 2.3 & 23 & 0.85 & 0.105 & 0.22 & 1.0 & 1.0 \\
Mirrored triplet & 2000 & 1.5 & 21 & 0.65 & 0.13 & 0.17 & 1.0 & 1.0 \\
Mirrored quartet & 2000 & 1.3 & 20 & 0.55 & 0.13 & 0.13 & 1.0 & 1.0 \\
\hline
Quartet+Dipole & 1000 & 1.6 & 5.3 & 0.85 & 0.344 & 0.342 \\
\end{tabular}
\end{ruledtabular}
\end{table*}

The drift lengths between the PMQs selected for these triplets can be changed to focus different energies in the same manner as studied for the doublet in the previous section. Extending the total system length to 1~m, the PMQs from the 560~mm triplet can be arranged to focus up to 117~MeV proton beams, and its system parameters are listed in Table~\ref{tab:SysPrm} as Triplet @ 117 MeV. This energy is significantly lower than the doublet discussed in Sec.~\ref{sbsec:dblt}. If an application requires a wide range of energies including up to > 200 MeV to be focused within the limited space, the doublet can be the best option, with the mentioned disadvantage of an asymmetric focus.

\subsubsection{Improving magnification symmetry of the transport system with a quartet configuration}
\begin{figure}[!h]
\includegraphics[width = .45\textwidth]{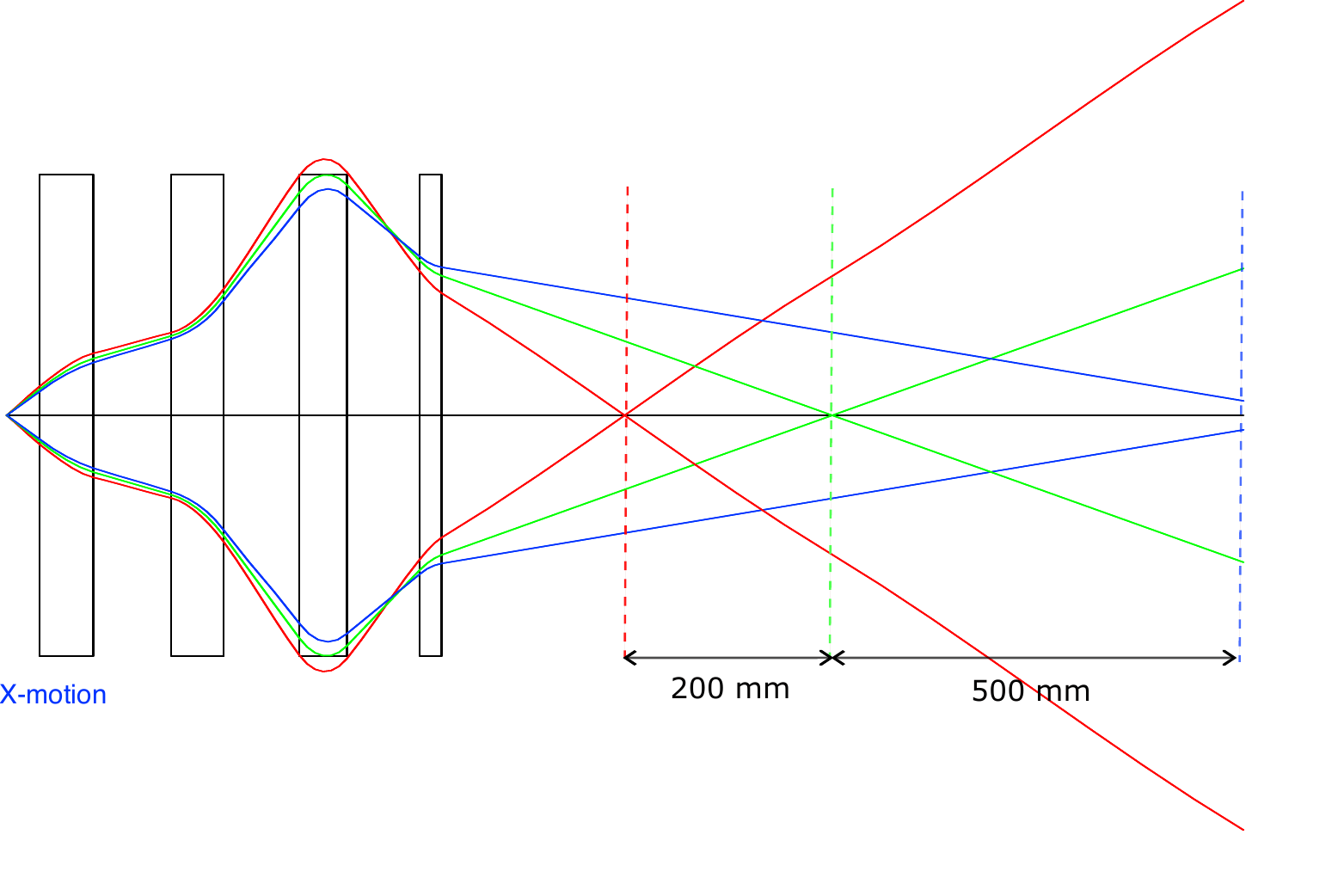}
\caption{\label{fig:qrtt} Illustration of the 1.0~m quartet for $E_0 = 30$~MeV proton beam. The red, green and blue lines show the envelopes of $0.85~E_0, E_0$, and $1.15~E_0$ proton beams. The half angle acceptances of the traces are 28.5~mrad for x and 32.5~mrad for y planes. Indicated by the dashed red (blue) line represent the optimal sample locations for delivering a lower (higher) average energy beam.}
\end{figure}

By adding a fourth focusing magnet, the magnification of the system, and thus the divergence of the beam at the output plane, can be made symmetric. The 1~m quartet design with symmetric magnification and divergence ($M_{12} = M_{34} = 0$, $M_{11} = M_{33}$, $M_{22} = M_{44}$) is illustrated in Fig.~\ref{fig:qrtt}. The system parameters and performance parameters are shown in Table~\ref{tab:SysPrm} and Table~\ref{tab:cnfg}. The PMQs from the shortest quartet design (610~mm) can be arranged to focus 92 MeV proton beam with a 1~m total system length, and its system parameters are listed in Table~\ref{tab:SysPrm} as Quartet @ 92 MeV.

One can see from Table~\ref{tab:cnfg} that the symmetry was further improved by the quartet designs for both shortest and 1~m cases, while other parameters were found to be comparable. While the energy acceptance, collection efficiency $\eta$, and energy acceptance $\delta E$ of the quartet configurations that can be focused within 1~m was comparable to the triplet design, the added beam symmetry at the output can be beneficial. For the doublet and triplet configurations, the lower (higher) energy beam focus location in x and y planes are far from each other along z due to chromatic aberrations. In the quartet design, one of the leading terms for chromatic aberrations can be balanced between different planes along z, such as $M_{126} = M_{346}$ or $M_{226} = M_{446}$. For the 1~m quartet design, $M_{126} = M_{346}$ was also added to the requirements. With this, the quartet as a whole can focus more symmetrically, similar to an active plasma lens~\cite{vanTilborg_2015}. 

\begin{figure}[!b]
\includegraphics[width = .45\textwidth]{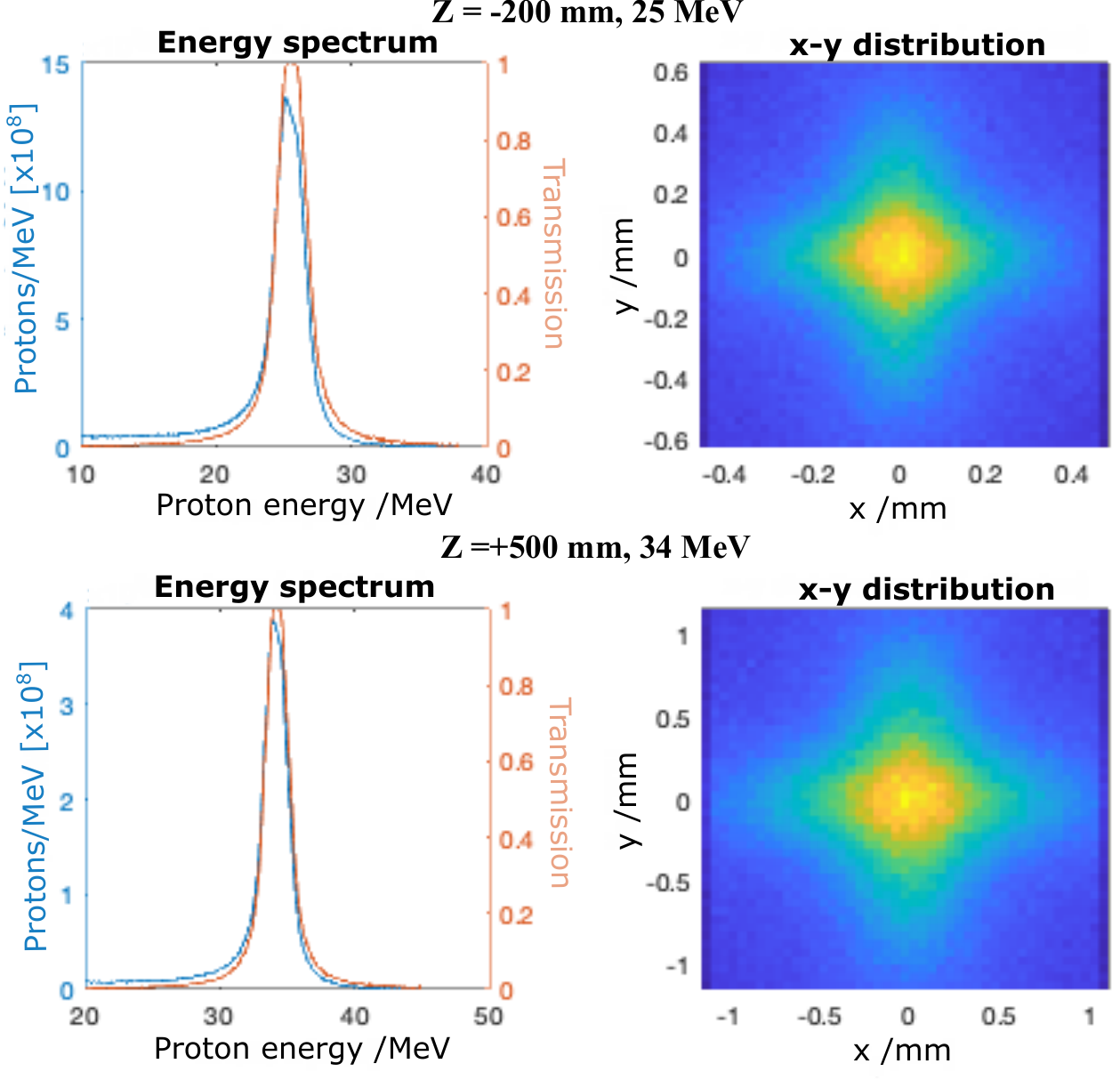}
\caption{\label{fig:qqqq_e} Simulated beam spectrum (left) and profile (right) at different target locations $z = -200$ mm (top) and $z = +500$ mm (bottom) relative to the 30~MeV focal location for the quartet design shown in Fig \ref{fig:qrtt}. This shows the target location can be used to tune the system to deliver different beam energies.}
\end{figure}

After other configurations are constructed, the longitudinal locations for the PMQs are the only free parameters that could be modified to vary the energy spectrum. For the quartet, it could more simply be done by varying the location of the sample to match the different focal position for a desired peak energy in the spectrum. This creates the effect of shifting the beam spectrum hitting the sample as off-energy particles would be defocused at these locations, lowering their relative intensity. The transverse profile and energy spectrum of the beam described in section B, transported with the 1~m quartet for different locations along z after the final PMQ, i.e, z = -200 (corresponding to a focused energy of 25.1 MeV) and z=+500~mm (corresponding to a focused energy of 34 MeV) with respect to the nominal output plane for a focused 30 MeV beam, are shown in Fig.~\ref{fig:qqqq_e}, and performance parameters are listed in Table~\ref{tab:cnfg} as "Quartet @ 26~MeV" and "Quartet 34~MeV". Depending on the application, a rough spectral scan can be performed by moving the target location rather than moving each PMQs. 

\subsubsection{Mirrored design for 1 to 1 imaging}
With the shortest designs, the magnification factor of the Quartet described in the previous section is close to 1 to 1 comparing the input beam with the transported beam at the output plane, and the 1~m Quartet design has a larger magnification as show in Table~\ref{tab:cnfg}. When an application requires 1 to 1 imaging or a high proton fluence on the sample (keeping the first magnet close to the source for high collection efficiency) while the sample location is relatively far from the source, one can employ a mirrored design. This is simply repeating the same transport in a mirrored manner. Furthermore, an iris or pinhole can be placed at the first focus location to provide a narrower energy spread. Mirroring the system creates a new symmetry which can compensate for some higher order chromatic and spherical terms in the transfer map as shown in reference \cite{Symmetry}. Shown in Table~\ref{tab:cnfg} is the performance estimate for the mirrored 1.0m doublet (total length 2.0~m) with an iris opening matching the FWHM of the beam size. One can see that for all three configurations the mirrored design can provide similar peak proton densities and energy spreads. The doublet has a slightly higher peak density than triplet and quartet, while the quartet provides the lowest energy spread at the output plane. It is note-worthy that the mirrored design is 1-to-1 not only for imaging ($M_{11} = M_{33} = 1$) but also for divergence ($M_{22} = M_{44} = 1$). If an application requires a high peak density relatively far away from the source ($\geq 3$ m), the mirrored design becomes a reasonable option.

\subsection{Controlling energy spread}
The energy select-ability of PMQ-based focusing systems has been discussed in sections above, as indicated by the energy spread $\delta E$. This energy selection scheme is not perfect, however, as a large number of the higher and lower energy particles can still make it through even if that constitutes a smaller fraction of the original particle numbers. As shown in Fig.~\ref{fig:qqqq_e} top left, the energy spectrum can have long low-energy tail. Some application may require complete removal of those high- and low-energy tails, and/or a narrower energy spread, e.g. for biological irradiation. Particles with different energies have different penetration depths and could lead to radiation damage at locations different from the targeted depth in the sample.

More precise energy selection can be realized by using dipole magnets in the transport beamline combined with a slit. By varying the slit location and size along the energy dispersion axis, one can control the energy and energy spread. However, this system introduces an energy dependent propagation angle, which may introduce inconvenience. Here we consider a pair of identical dipole magnets (dog-leg) and an optional slit. This way the propagation angle remains parallel to the input angle, although different energies of the beam now propagate along different axes with a lateral offset from the original propagation axis. 

\begin{figure}[!t]
\includegraphics[width = .45\textwidth]{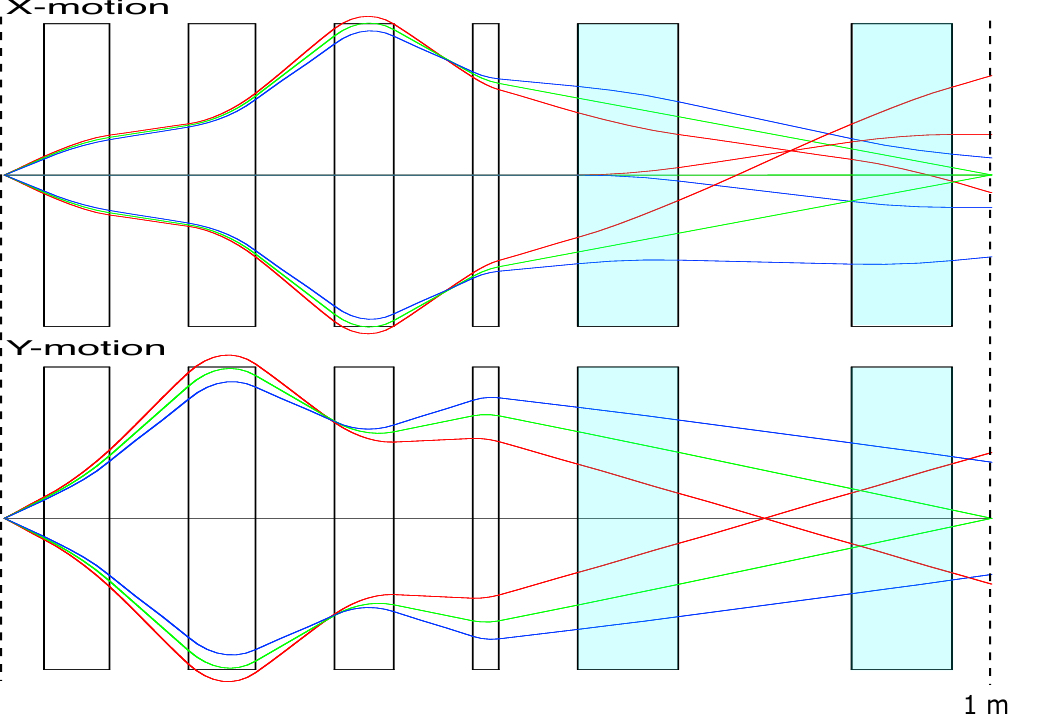}
\caption{\label{fig:qrttDl} Illustration of the 1.0~m quartet with two dipole magnets from Sec. \ref{sec:PMD} with magnetic dispersion along the x axis (Dipole field oriented along y) for $E_0 = 30$~MeV proton beam. The red, green and blue lines show the envelopes of $0.85~E_0, E_0$, and $1.15~E_0$ proton beams. The half angle of the traces are 28.5~mrad for x and 32.5~mrad for y planes. The dipole magnets are shown in blue.}
\end{figure}

Illustrated in Fig.~\ref{fig:qrttDl} is a 1.0~m quartet system with two 101.6~mm-long PMDs, its system and performance parameters and are shown in Table~\ref{tab:SysPrm} and \ref{tab:cnfg}, respectively. The system was designed to satisfy $M_{12} = M_{34} = 0$, $M_{11} = M_{33}$, $M_{22} = M_{44}$, and is nearly identical to the 1m quartet design. Slight modifications in the magnet lengths are to compensate weak focusing forces from the PMDs. One can see that the energy spread was reduced from 1.05 to 0.85~MeV. Note that this is achieved without a slit, but by only taking into account the beam portion within the $\pm 1$ FWHM beam size. The energy spread can be improved further by limiting the beam in the dispersion plane with a slit. The slight decrease in the peak density is from the removal of lower/higher energy particles contributions. From Fig.~\ref{fig:qrttDl}, the defocused 0.85~E and 1.15~E beams are still partially overlapped. The energy spread can be improved by having longer drift length between two PMDs or using longer PMDs. Here, the total length was limited to 1~m to show what can be reasonably achieved within 1~m. Since the last drift lengths for the 1~m doublet and 1~m triplet are much longer, a narrower energy spread within the same length can be realized with only two dipoles. If an application demands a small energy spread while being compact, the doublet or triplet can become better options. Note that this design can also be mirrored to provide 1-to-1 imaging.

\begin{figure}[t]
\includegraphics[width = .4\textwidth]{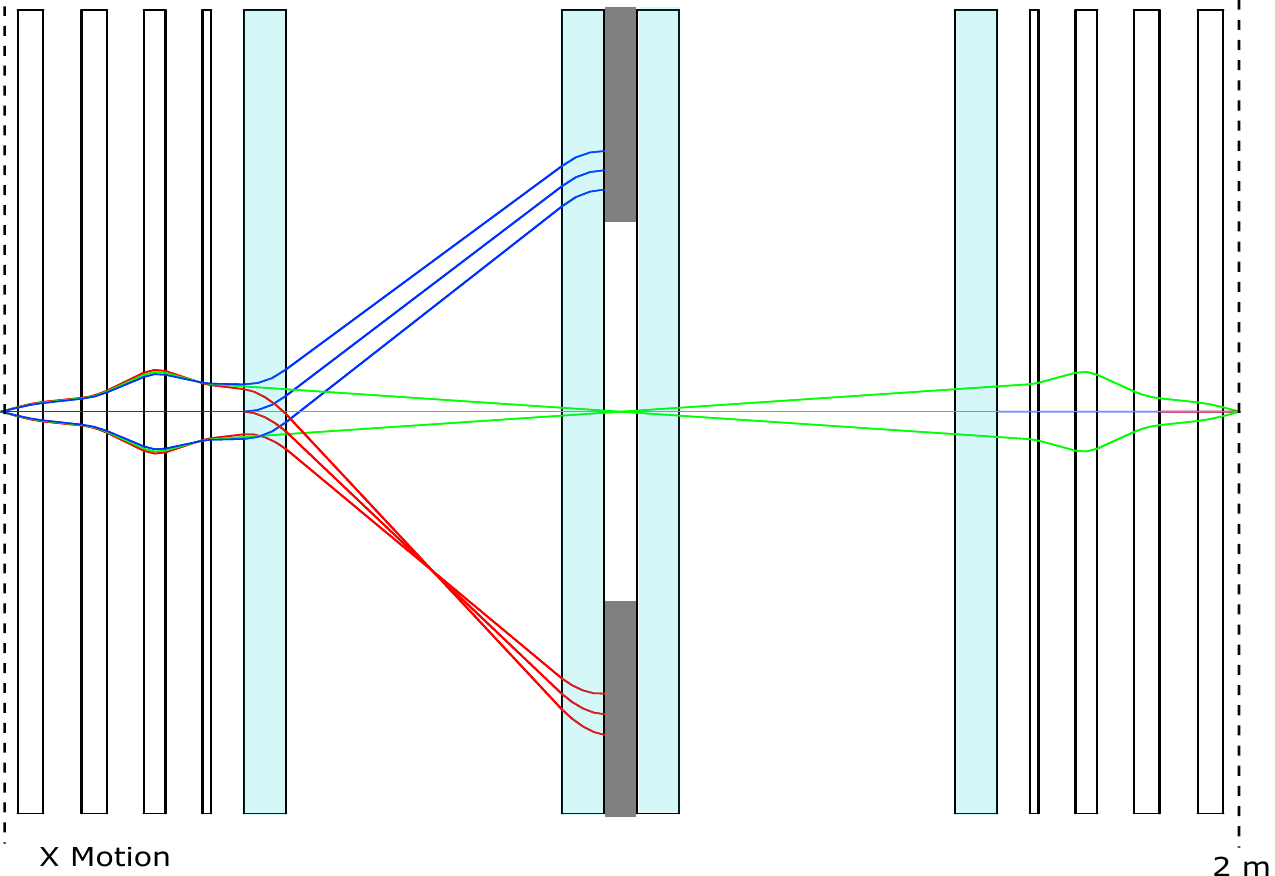}

\caption{\label{fig:mirrored_ESS}Mirrored quartet simple particle tracking. Higher energies (blue) and lower energies (red) are blocked by slit. Energy can be selected by changing magnet and slit (gray).}
\end{figure}

The energy spread can be better controlled using an Energy Selection System (ESS) as outlined in \cite{Romano_2016}. It consists of a set of four dipoles placed between the mirrored quadrupoles. Similar to a chicane bunch compressor or fragment separator, the dipoles are arranged to kick the beam off and back onto the longitudinal axis. Particles of different energies will take different paths through the system due to the velocity dependence of the Lorentz force and a beam block, or slit, can be arranged to only allow particles in a desired energy range to pass through. A diagram of this design for the quartet is shown in Fig~\ref{fig:mirrored_ESS}.

\begin{table*}[t]
\small
\caption{\label{tab:energy_tune}%
Performance of energy tuning doublet and triplet designs }
\begin{ruledtabular}
\begin{tabular}{ccccccc}
Energy & $\eta$ & $n_{1pk}$ & $\delta E$ & $\sigma_{x1}$ & $\sigma_{y1}$& Total Length\\
~[MeV] & [$\%$] & [10$^9$/mm$^2$] & [MeV] & [mm] & [mm] & [mm]\\ 
\hline
\multicolumn{7}{c}{\textrm{Doublet}}\\
\hline
30&2.61&2.04&1.3&1.98&0.21&1000\\
35&1.58&2.18&0.8&1.56&0.09&783\\
40&1.42&1.19&0.9&1.59&0.09&822\\
45&0.84&0.49&0.9&1.92&0.09&959\\
\hline
\multicolumn{7}{c}{\textrm{Triplet}}\\
\hline
30&1.56&5.15&1.2&0.30&0.48&1000\\
35&1.12&2.73&0.9&0.33&0.33&1000\\
40&0.96&1.34&1.0&0.33&0.39&1000\\
45&0.77&0.83&0.8&0.24&0.39&1000\\
\end{tabular}
\end{ruledtabular}
\end{table*}

\subsection{10 MeV Collimator for radiobiology} \label{sec:10MeVCol}
The previous work establishes the tools necessary to efficiently design beamlines for a variety of applications in laser-driven ion acceleration. As an example, we present a beamline design suitable for radiobiology, particularly for high-dose-rate irradiation studies.

In radiobiological applications, uniform irradiation of samples is required to ensure accurate dose delivery. This necessitates a shift from point focusing to beam collimation to ensure a uniform beam spot, in this case, 7~mm in diameter. For thin biological samples (e.g. 2D cell cultures, mouse ears), the beam energy is able to be relaxed (10 MeV) to lower the necessary focusing strength of the transport while still achieving full penetration in the sample.  The flexibility of the transport simulation tools discussed above enabled the development of a tailored design to meet these specific requirements.

The design consists of a pair of PMQs and a single dipole with specifications shown in \ref{tab:10MeVColl}. This design was implemented and used in an experiment at the iP2 beamline \cite{dechant_2024} where the team planned to irradiate \textit{in vivo} samples to study ultra-high dose rate radiobiology, an application well-suited to LD ion beams due their high intensity. The results of this study and performance of the transport are pending and will be addressed in a future publication.


\begin{table}[t]
\caption{\label{tab:10MeVColl}
Magnet specifications for the 10 MeV collimator scheme installed in iP2}
\begin{tabular}{c|cccccc}
\hline\hline
& R$_{in}$ & R$_{out}$ & B' & L$_{eff}$ & B’ x Z & B$_{tip}$  \\
& [mm] & [mm] & [T/m] & [mm] & [T] & [T] \\\hline
M1 & 5  & 15 & 250 & 50 & 12.5 & 1.25  \\
M2 & 15 & 45 & 67  & 50 & 3.35 & 1.005\\ \hline\hline
\end{tabular}
\end{table}

\section{Conclusion}
We studied the collection and focusing of laser-driven proton beams with compact, permanent magnet-based beam optics. A wide range of compact designs were explored in this study, each with their own advantages and disadvantages. The doublet delivers the highest transport efficiency and can have the most compact footprint, but it suffers from an asymmetric beam spot at the output plane (sample site). The quartet produces the most symmetric beam spot at the sample site with a more uniform energy distribution, but is more costly than the other designs, because it requires more magnets, occupies more real estate in the experimental chamber, and is more sensitive to alignment errors. Mirroring each configuration increases the proton spot quality and, thus, proton peak density at the sample site because of the compensation of aberrations, and allows for energy selection, either with just a slit or the energy selection system. This also doubles the number of magnets and total required length of the system. The choice of design thus depends on the application of the transport. We used the here established transport simulation tools to design a 10 MeV collimation system for the first \textit{in vivo} normal tissue irradiation with laser driven protons

There are many areas of future study. The effect of small perturbations or misalignment of the magnets on performance should be investigated along with other practical considerations like fringe fields. While high order aberrations higher than 3rd order are unlikely to create a significant difference, a more thorough study would be worth pursuing. All the configurations still have rather significant chromatic aberrations which could affect the practicality of them for biological applications. A further study using higher order magnets (e.g. sextupoles and octupoles) to control for these effects should be done. 

The work presented here is significant to the community of laser-driven ion beam researchers, as well as those involved in accelerator technology and high-energy physics. The designs developed in this study will enable more efficient ion collection and transport for LDIA applications ranging from cancer therapy to material science. Notably, the study provides a pathway to addressing the long-standing challenge of managing the large divergence and energy spread in a compact environment, which is a critical barrier to advancing LDIA technology.

\begin{acknowledgments}
We would like to thank Jian-Hua Mao, Jamie Inman, Sven Steinke, Thomas Schenkel, Aodhan McIlvenny, Brendan Stassel, Sahel Hakimi for their contributions. J. De Chant would like to thank Steve Lidia for guidance and sharing his expertise. The work was supported by Laboratory Directed Research and Development (LDRD) funding from Lawrence Berkeley National Laboratory provided by the Director, and the U.S. Department of Energy Office of Science, Offices of Fusion Energy Sciences, High Energy Physics and LaserNetUS under Contract No. DE-AC02-05CH11231.
\end{acknowledgments}




\bibliography{citations.bib}

\end{document}